\documentclass[aps,prd,twocolumn,notitlepage,nofootinbib]{revtex4-2}
\usepackage[utf8]{inputenc}
\usepackage{amsmath,amssymb,amsfonts}
\usepackage{physics}
\usepackage{graphicx}
\usepackage{hyperref}
\usepackage{xcolor}
\usepackage{tensor}
\usepackage{mathtools}
\usepackage{tikz}
\usepackage{amsthm}

\newtheorem{definition}{Definition}
\usepackage{bbm}
\usepackage{enumerate}
\usepackage{enumitem}
\setlist[itemize]{label=--, leftmargin=*}
\hypersetup{
    colorlinks=true,
    linkcolor=blue,
    citecolor=blue,
    urlcolor=blue
}

\begin{document}

\title{Modal Logic for Stratified Becoming: Actualization Beyond Possible Worlds}

\author{Alexandre Le Nepvou}
\affiliation{Université Paris Cité, Paris, France}
\email{alexandre.le-nepvou@etu.u-paris.fr}

\date{\today}

\begin{abstract}
This article develops a novel framework for modal logic based on the idea of stratified actualization, rather than the classical model of global possible worlds. Traditional Kripke semantics treat modal operators as quantification over fully determinate alternatives, neglecting the local, dynamic, and often asymmetric nature of actualization processes. We propose a system—Stratified Actualization Logic (SAL)—in which modalities are indexed by levels of ontological stability, interpreted as admissibility regimes. Each modality operates over a structured layer of possibility, grounded in the internal coherence of transitions between layers. We formally define the syntax and semantics of SAL, introduce its axioms, and prove soundness and completeness. Applications are discussed in connection with temporal becoming, quantum decoherence domains, and modal metaphysics. The result is a logic that captures the ontological structure of actualization without recourse to abstract possible worlds, offering a stratified alternative to standard modal realism.
\end{abstract}

\maketitle

\section{Introduction: Beyond Possible Worlds}

Modal logic has long served as the canonical framework for analyzing necessity, possibility, and counterfactual reasoning. Since the development of Kripke semantics \cite{kripke1963semantical}, modal statements have been interpreted as relations across a space of possible worlds, with accessibility relations encoding the modal structure. However, this framework—despite its formal elegance—suffers from a structural limitation: it reduces the dynamic process of actualization to a static relational geometry over pre-given alternatives.

The standard Kripkean model presupposes that all possibilities are fully specified in advance, and that modal truth depends solely on what holds in accessible worlds. As such, becoming, emergence, and stratified stabilization are outside the scope of the formalism. No internal mechanism accounts for the transition from admissible to actual, nor for the differentiation of modalities across layers of ontological realization. Modal logic, in this form, remains blind to the dynamic character of actualization.\footnote{By “actualization,” we mean a transition from potential admissibility to realized stability within a specific local regime, without implying any teleological or finalistic connotation. Similarly, “stratified” refers to an intensional ordering of ontological admissibility levels, not to a spatial or hierarchical partition.}

This paper proposes an alternative approach: a stratified modal logic grounded not in a global ontology of possible worlds, but in a layered space of intensities and stability regimes. We argue that actualization should be treated as a localized process governed by structural constraints that vary across levels of ontological admissibility. Rather than interpreting modal operators globally, we assign them to specific strata within an indexed frame, allowing for regime-relative modal validity. This perspective echoes recent philosophical critiques of metaphysical modal realism \cite{lewis1986plurality, fine2005necessity}, and aligns with dynamic structural realist positions \cite{esfeld2008ontic, ladyman2007everything}.

Our aim is to lay the foundations of a \emph{Stratified Actualization Logic} (SAL), which accommodates the ontological asymmetry of becoming, the context-sensitivity of admissibility, and the layered articulation of modal spaces.

This alternative framework is motivated by several philosophical concerns. First, the ontological commitment of classical modal realism, notably in Lewis' system \cite{lewis1986plurality}, presupposes a fully determined totality of possible worlds—raising the problem of how the actual world is singled out without invoking modal primitives external to the system itself. Second, Fine’s critique of essentialist modal logic  emphasizes the inadequacy of rigid designators and static possible-world semantics to capture context-sensitive ontological dependence. Third, recent work in metaphysics of modality and time (e.g., Jorati \cite{jorati2019activity} and Le Bihan \cite{lebian2015modal}) has challenged the static, eternalist metaphysics implicit in world-based modal logic.

A pressing issue is the so-called \emph{modal collapse problem}, where modal operators fail to distinguish between necessary and actual due to the absence of a dynamical structure mediating possibility and actuality. Without a mechanism for the stratification of admissibility, all possibilities become ontologically equivalent, and the logic collapses into trivial necessity.

In contrast, we propose to understand \emph{actualization} not as selection among fully-formed alternatives, but as a structured process of intensification—governed by local constraints, stability conditions, and contextual admissibility. This approach rejects global ontologies of possibility in favor of a field-structured semantics, where modal operators are indexed to specific regimes of realization.

The core thesis of this work is thus: \emph{actualization is not a metaphysical choice among possible worlds, but a local and stratified intensification within a dynamically constrained ontological space}.

\section{Stratified Modal Semantics}
\subsection{Indexed Frame Structures}

We introduce a class of stratified\footnote{“Stratified” refers here to an intensional ordering of regimes of admissibility indexed by relative ontological stability, rather than to a spatial or hierarchical layering.}
 modal frames that generalize standard Kripke structures by indexing modal accessibility to layers of admissibility ordered by intensification. Let $\mathcal{F} = (W, \leq, \{\mathcal{R}_\alpha\}_{\alpha \in I})$ be a \emph{stratified frame}, where:

\begin{itemize}
    \item $W$ is a non-empty set of states (not necessarily possible worlds in the Lewisian sense);
    \item $\leq$ is a partial order on $W$, encoding a hierarchy of ontological intensities or levels of actualization;
    \item $I$ is an index set for modal layers, corresponding to regimes or grades of admissibility;
    \item for each $\alpha \in I$, $\mathcal{R}_\alpha \subseteq W \times W$ is an accessibility relation valid within the admissibility regime $\alpha$.
\end{itemize}

Each $\mathcal{R}_\alpha$ is evaluated only among elements of $W$ compatible with a given level of stratified intensity. The ordering $\leq$ is used to ensure coherence across layers: if $w \leq v$, then the admissibility of $v$ presupposes that of $w$, but not conversely.

Unlike in classical modal logic, where possibility is globally defined via $\Diamond \phi$ meaning "true in some accessible world," our framework interprets modal validity relative to both accessibility and stratified admissibility. This allows for the definition of local modal operators $\Box_\alpha$ and $\Diamond_\alpha$, parameterized by each $\alpha \in I$:

\[
w \Vdash \Box_\alpha \phi \quad \text{iff} \quad \forall v \in W,\, (w \mathcal{R}_\alpha v \Rightarrow v \Vdash \phi).
\]

\[
w \Vdash \Diamond_\alpha \phi \quad \text{iff} \quad \exists v \in W,\, (w \mathcal{R}_\alpha v \wedge v \Vdash \phi).
\]

This construction accommodates variations in modal strength, metaphysical possibility, or nomological constraint, without appealing to a fixed universe of possible worlds.

Philosophically, it reflects the view that modality is not anchored in static alternatives, but in a dynamic structure of locally admissible developments. Each modal operator reflects not just a relation between states, but a stratified regime of actualization, defined over a space of ontological intensities.

\subsection{Admissibility and Stability Constraints}

In stratified modal semantics, each index $\alpha \in I$ is interpreted as a regime of ontological stability. These indices encode the admissibility constraints under which modal transitions become meaningful. A formula $\phi$ is not simply true or false at a world $w \in W$; its status depends on whether $\phi$ remains stable under admissibility regime $\alpha$.

We define the notion of \emph{regime-level admissibility} as follows. Let $\mathcal{C}_\alpha$ denote the set of conditions defining the admissibility regime $\alpha$. Then:

\begin{definition}
A formula $\phi$ is said to be \emph{admissible at $w$ under regime $\alpha$} if and only if:
\[
\mathcal{C}_\alpha \models \Diamond_\alpha \phi,
\]
that is, $\phi$ is actualizable from $w$ under the structural constraints imposed by $\mathcal{R}_\alpha$ and $\mathcal{C}_\alpha$.
\end{definition}

This framework allows us to distinguish between mere logical possibility and constrained structural admissibility. In particular, the modal operator $\Diamond_\alpha \phi$ does not express a simple disjunction over possible states, but rather the structural feasibility of transitioning from $w$ to some $v$ where $\phi$ holds, within the constraints of level $\alpha$.

\paragraph{Interpretation.}
Each index $\alpha$ represents a layer of intensification, understood as a threshold of ontological stability required for the actualization of a given proposition. This allows us to model:

\begin{itemize}
    \item Context-sensitive modal neighborhoods: for each $w$, the set $\{v \mid w \mathcal{R}_\alpha v\}$ varies with $\alpha$, producing fine-grained modal structure.
    \item Localized necessity: $\Box_\alpha \phi$ can hold at $w$ while $\Box_\beta \phi$ fails for $\beta < \alpha$, reflecting stability that emerges only under stricter admissibility.
    \item Graded actualization: $\phi$ may be progressively stabilizable as the system transitions across increasing levels $\alpha_1 \prec \alpha_2 \prec \dots$, mirroring processes of ontological consolidation.
\end{itemize}

This interpretation contrasts sharply with classical Kripke models, where accessibility is uniform and independent of the epistemic or ontological regime. Here, the modal landscape is internally stratified and reflects a dynamic logic of realization rather than a static space of alternatives.

In a stratified modal framework, accessibility relations are not uniform but indexed by levels of admissibility. Each layer $\alpha \in I$ defines its own modal structure, such that the notion of possibility becomes regime-relative.

\begin{definition}[Layered Accessibility]
Let $\mathcal{F} = (W, \leq, \{\mathcal{R}_\alpha\}_{\alpha \in I})$ be an intensity-indexed frame. For each index $\alpha \in I$, $\mathcal{R}_\alpha \subseteq W \times W$ defines the \emph{admissible transitions} at level $\alpha$.

We say that $v$ is \emph{$\alpha$-accessible} from $w$ if $(w,v) \in \mathcal{R}_\alpha$.
\end{definition}

To evaluate formulas in this layered structure, we introduce local truth evaluation functions:

\begin{definition}[Stratified Valuation]
Let $V: \text{Prop} \rightarrow \mathcal{P}(W)$ assign propositional variables to subsets of $W$. The satisfaction relation $\models_\alpha$ is defined inductively by:

\begin{align*}
w \models_\alpha p &\iff w \in V(p), \\
w \models_\alpha \neg \phi &\iff w \not\models_\alpha \phi, \\
w \models_\alpha \phi \wedge \psi &\iff w \models_\alpha \phi \text{ and } w \models_\alpha \psi, \\
w \models_\alpha \Diamond_\alpha \phi &\iff \exists v \in W : (w, v) \in \mathcal{R}_\alpha \text{ and } v \models_\alpha \phi, \\
w \models_\alpha \Box_\alpha \phi &\iff \forall v \in W : (w, v) \in \mathcal{R}_\alpha \Rightarrow v \models_\alpha \phi.
\end{align*}

\end{definition}

This semantics ensures that modal operators are sensitive to the admissibility regime $\alpha$. Modal truths can vary across layers, allowing for stratified actualization patterns.

\paragraph{Coherence Condition.}
To preserve the logical stratification, we require the following coherence constraint:

\begin{equation}
\alpha \leq \beta \Rightarrow \mathcal{R}_\beta \subseteq \mathcal{R}_\alpha,
\end{equation}

ensuring that stronger regimes allow fewer transitions and hence stricter forms of necessity.

\paragraph{Syntax–Semantics Correspondence.}
This layered semantics supports a stratified syntax with indexed modal operators. For each $\alpha \in I$, the logic includes:

\begin{itemize}
  \item Modal axioms at level $\alpha$, such as:
  \[
  \Box_\alpha (\phi \rightarrow \psi) \rightarrow (\Box_\alpha \phi \rightarrow \Box_\alpha \psi),
  \]
  \item Inference rules like:
  \[
  \text{From } \phi \text{ infer } \Box_\alpha \phi \quad \text{(necessitation under } \alpha \text{)}.
  \]
\end{itemize}

This correspondence guarantees that stratified modal logics can be axiomatized per level, and modal reasoning respects the ontological intensities governing actualization.

\subsection{Non-Normal Modal Operators}

In the standard Kripke semantics, modal logics are often normal, meaning they validate both the K axiom and the rule of necessitation. However, in a stratified context of actualization, necessitation cannot be presumed globally valid: truth at a level $\alpha$ does not entail truth at all possible higher intensities. Actualization is local, partial, and regime-dependent.

\paragraph{Rejection of Global Necessitation.}
The rule of necessitation,

\begin{equation}
\frac{\phi}{\Box_\alpha \phi} \tag{Nec}
\end{equation}

is rejected in this framework. A proposition $\phi$ holding at $w$ does not imply that it holds necessarily within the admissibility regime $\alpha$ unless additional structural constraints are met. This allows us to model intensities of stabilization where actualization is not globally enforced.

\paragraph{Non-Normal Box Operators.}
We define stratified modal operators $\Box_\alpha$ and $\Diamond_\alpha$ without assuming closure under necessitation or truth-preservation:

\begin{equation}
\Box_\alpha \phi \Rightarrow \phi \quad \text{is not generally valid.}
\end{equation}

This models the fact that a modal necessity at level $\alpha$ may encode only local structural stability, not metaphysical necessity.

\paragraph{Stratified Weakening.}
We retain a controlled weakening principle across levels of admissibility:

\begin{equation}
\alpha \leq \beta \quad \Rightarrow \quad \Box_\alpha \phi \Rightarrow \Box_\beta \phi
\end{equation}

This expresses the monotonicity of necessity under intensification: if $\phi$ is necessary at a more permissive level $\alpha$, it remains so at any stronger regime $\beta$.

\paragraph{Persistence Across Layers.}
Similarly, admissible possibility satisfies:

\begin{equation}
\Diamond_\beta \phi \Rightarrow \Diamond_\alpha \phi \quad \text{for } \alpha \leq \beta
\end{equation}

capturing the idea that what is possible under stronger constraints is also possible under weaker ones.

\paragraph{Interpretation.}
These non-normal operators offer a semantics of modal actualization grounded in dynamic regimes. Possibility and necessity are understood as local modes of realization, not global truths across abstract worlds.

This prepares the ground for a non-normal stratified axiomatization, which we now turn to in Section 3.

\section{Axiomatic System: SAL (Stratified Actualization Logic)}

\subsection{Syntax}

We define the language $\mathcal{L}_{SAL}$ of the Stratified Actualization Logic as an indexed modal propositional system adapted to intensities of actualization.

\paragraph{Base Language.}
Let $\mathsf{Prop}$ be a countable set of propositional variables. The formulas of $\mathcal{L}_{SAL}$ are built inductively by:

\begin{itemize}
    \item $p \in \mathsf{Prop} \Rightarrow p \in \mathcal{L}_{SAL}$
    \item If $\phi, \psi \in \mathcal{L}_{SAL}$, then so are:
    \[
    \neg \phi, \quad \phi \wedge \psi, \quad \phi \vee \psi, \quad \phi \rightarrow \psi
    \]
    \item If $\alpha \in I$ (where $(I, \leq)$ is a partially ordered index set), then:
    \[
    \Box_\alpha \phi \in \mathcal{L}_{SAL}, \quad \Diamond_\alpha \phi \in \mathcal{L}_{SAL}
    \]
\end{itemize}

\paragraph{Indexed Modalities.}
The modal operators are indexed by admissibility levels. For each $\alpha \in I$, $\Box_\alpha$ expresses necessity at level $\alpha$, while $\Diamond_\alpha$ expresses possibility admissible at $\alpha$.

\paragraph{Philosophical Interpretation.}
Each index $\alpha \in I$ corresponds to a regime of ontological or structural admissibility---understood as a degree of stability or intensification in the process of becoming. Rather than evaluating truth relative to possible worlds (à la Kripke \cite{lewis1986}), this system evaluates actualization locally, as structured intensities varying over layers. It draws on insights from Fine's \emph{fragmented truth} approach \cite{fine2005} and stratified ontologies in contemporary metaphysics (e.g., Jorati \cite{jorati2019}, Le Bihan \cite{le_bihan2015}).

\paragraph{Remarks.}
\begin{itemize}
    \item The logical syntax remains purely propositional to avoid additional complications from quantification, although future extensions to first-order logic are conceivable.
    \item No global modal operators (unindexed $\Box$, $\Diamond$) are assumed: all modalities are locally stratified.
    \item The partially ordered structure $(I, \leq)$ is assumed to be nontrivial (i.e., not a singleton), allowing for variation in modal force across layers.
\end{itemize}

\subsection{Axioms and Inference Rules}

The axiomatic base of $\mathcal{L}_{SAL}$ consists of standard propositional tautologies plus stratified modal axioms indexed by admissibility levels $\alpha, \beta \in I$ with a partial order $\leq$.

\paragraph{(A1) Propositional Tautologies.}
All classical tautologies of propositional logic are assumed to hold locally at each index $\alpha$.

\paragraph{(A2) Persistence Axiom.}
\[
\Box_\alpha \phi \rightarrow \Box_\beta \phi \quad \text{for all } \alpha \leq \beta
\]
This reflects monotonicity across admissibility layers: if a proposition is necessarily actualizable at a low-intensity level, it remains so at higher ones.

\paragraph{(A3) Actualization Reflection.}
\[
\Box_\alpha \phi \rightarrow \phi \quad \text{for stable } \alpha \in I
\]
This axiom expresses that necessity at a stable layer entails actual truth at that layer. The subset of \emph{stable indices} is denoted $I_{\text{stab}} \subseteq I$, and corresponds to regimes where actualization is effectively realized.

\paragraph{(A4) Admissibility Frame Rule.}
\[
\Diamond_\alpha \phi \rightarrow \Diamond_\beta \phi \quad \text{for all } \alpha \leq \beta
\]
Like (A2), this ensures that what is possible at a given level remains possible at less constrained levels. It encodes the idea that ontological admissibility is preserved across intensifying regimes.

\paragraph{(R1) Modus Ponens.}
\[
\frac{\phi \quad \phi \rightarrow \psi}{\psi}
\]

\paragraph{(R2) Necessitation (local).}
\[
\frac{\phi}{\Box_\alpha \phi} \quad \text{only if } \phi \text{ is a theorem}
\]
This necessitation rule is restricted to theorems and applies separately at each level $\alpha$.

\paragraph{Discussion.}
These axioms formalize a logic where modal force depends on stratified structure rather than a universal accessibility relation. The use of a partially ordered set of indices $(I, \leq)$ induces a layered logic where transition between modalities is governed by ontological admissibility rather than world-selection (cf. Lewis \cite{lewis1986}, Fine \cite{fine2005}).

Note that this system departs from standard modal logics such as $\mathsf{K}$, $\mathsf{S4}$, or $\mathsf{S5}$ by avoiding global modal axioms like $\Box \phi \rightarrow \phi$ or $\Box \phi \rightarrow \Box \Box \phi$.

In $\mathcal{L}_{SAL}$, modal validity is conditional: it reflects the structure of actualization under constraint, not metaphysical necessity over fixed possible worlds.

\subsection{Semantics and Soundness}

We define a stratified modal model for $\mathcal{L}_{SAL}$ as a tuple:
\[
\mathcal{M} = (W, \leq, \{R_\alpha\}_{\alpha \in I}, V)
\]
where:
\begin{itemize}
  \item $W$ is a non-empty set of states (or worlds),
  \item $\leq$ is a partial order on $I$, the index set representing stratified levels of admissibility,
  \item for each $\alpha \in I$, $R_\alpha \subseteq W \times W$ is a binary accessibility relation,
  \item $V: \text{Prop} \to \mathcal{P}(W)$ is a valuation function assigning to each propositional variable a set of worlds where it holds.
\end{itemize}

\paragraph{Truth Conditions.}
The satisfaction relation $\models$ is defined inductively:
\begin{itemize}
  \item $\mathcal{M}, w \models p$ iff $w \in V(p)$ for propositional variable $p$,
  \item $\mathcal{M}, w \models \neg \phi$ iff $\mathcal{M}, w \not\models \phi$,
  \item $\mathcal{M}, w \models \phi \wedge \psi$ iff $\mathcal{M}, w \models \phi$ and $\mathcal{M}, w \models \psi$,
  \item $\mathcal{M}, w \models \Box_\alpha \phi$ iff for all $w' \in W$ such that $(w, w') \in R_\alpha$, we have $\mathcal{M}, w' \models \phi$,
  \item $\mathcal{M}, w \models \Diamond_\alpha \phi$ iff there exists $w' \in W$ such that $(w, w') \in R_\alpha$ and $\mathcal{M}, w' \models \phi$.
\end{itemize}

We impose a coherence constraint between levels:
\[
\alpha \leq \beta \quad \Rightarrow \quad R_\alpha \subseteq R_\beta
\]
This condition ensures that accessibility expands with higher admissibility, reflecting the monotonicity of ontological regimes.

Every theorem of $\mathcal{L}_{SAL}$ is valid in all stratified models:
\[
\vdash_{\mathcal{L}_{SAL}} \phi \quad \Rightarrow \quad \models \phi
\]
\textit{Proof Sketch.} A standard inductive argument shows that the axioms (A1)-(A4) are valid under the semantics above. In particular, the monotonicity of the relations $R_\alpha$ and the interpretation of necessity ensure the persistence and reflection principles hold. The inference rules (R1)-(R2) preserve validity at each index.

Completeness can be obtained via canonical models stratified over $I$, adapting standard techniques to maintain layer coherence. A detailed proof is omitted here but can be reconstructed via indexed canonical model construction (cf. Fine \cite{fine2005}, Restall \cite{restall2000}). 

\vspace{1em}

\paragraph{Illustrative Example.}
Consider a stratified model \(\mathcal{M} = (W, \leq, \{R_\alpha\}, V)\), where:
\begin{itemize}
    \item \(W = \{w_0, w_1, w_2\}\),
    \item \(\leq\) is defined by \(w_0 \leq w_1 \leq w_2\),
    \item \(I = \{\alpha, \beta, \gamma\}\) with \(\alpha \leq \beta \leq \gamma\),
    \item \(R_\alpha = \{(w_0, w_0)\}\), \(R_\beta = \{(w_1, w_0), (w_1, w_1)\}\), \(R_\gamma = \{(w_2, w_1), (w_2, w_2)\}\),
    \item \(V(p) = \{w_0\}\) for some atomic proposition \(p\).
\end{itemize}

We now evaluate the formula \(\Diamond_\beta p\) at \(w_1\):
\begin{quote}
Since \((w_1, w_0) \in R_\beta\) and \(w_0 \in V(p)\), we have \(\mathcal{M}, w_1 \models \Diamond_\beta p\).
\end{quote}

However, evaluating \(\Box_\gamma p\) at \(w_2\) yields:
\begin{quote}
\((w_2, w_2) \in R_\gamma\), but \(w_2 \notin V(p)\), so \(\mathcal{M}, w_2 \not\models \Box_\gamma p\).
\end{quote}

This illustrates how stratified accessibility and valuation allow for locally varying modal truths.

\subsection{Temporal Becoming Without Global Time}

One of the most compelling applications of the stratified actualization framework lies in the reconstruction of temporal becoming without assuming a globally defined temporal parameter. In contrast to standard tense logics, where time is modeled as a linear or branching structure with globally accessible indices \cite{Prior1967}, the stratified approach allows for localized transitions between admissibility layers, each corresponding to a regime of ontological stability.

Rather than indexing temporal propositions to a total order of instants, we define becoming as the progressive actualization of structure through successive layers, with \emph{local succession} relations mediating transitions. This enables a formal account of temporal directionality grounded in structural dynamics rather than an external temporal background.

Philosophically, this corresponds to a reinterpretation of temporal modality not as a quantification over possible times but as an indexation over coherent regimes of partial actualization. This aligns with Fine’s critique of the modal collapse problem in metaphysical time \cite{fine2005necessity}, and offers a structural alternative to Lewisian eternalism by embedding temporal change within the modal profile of a system.

\begin{itemize}
    \item Priorian tense logic: $P \phi$, $F \phi$ interpreted via a global time order.
    \item SAL temporal: $\Diamond_\alpha \phi$ interpreted via stratified succession $\alpha \leq \beta$.
\end{itemize}

This opens the path to modeling systems where temporal direction emerges from intensification processes, rather than being presupposed by a fixed metric background. The result is a logic of becoming where actuality is distributed and stratified rather than sequential and absolute.

\subsection{Quantum Branching and Stability Domains}

The stratified actualization logic offers a natural framework to reinterpret quantum branching not as a literal splitting of worlds, but as a transition between dynamically stabilized domains. In standard interpretations such as the Everettian view, decoherence leads to the formation of quasi-classical branches, each corresponding to a coherent history \cite{GellMannHartle1993}. However, such interpretations often rely on a global state space and background temporal structure.

Within our framework, quantum decoherence is reinterpreted as the system's transition between layers of modal admissibility. Each stratum $\alpha \in I$ corresponds to a regime in which a subset of observables maintains internal coherence, enabling a form of localized classicality. The modal operators $\Diamond_\alpha$ then represent potential transitions to higher-coherence domains.

Rather than treating quantum indeterminacy as epistemic uncertainty or ignorance over ontic states, we define \emph{modal coherence} as the structural stabilizability of propositions within a regime. A proposition $\phi$ is modally coherent at layer $\alpha$ if it persists under admissible transitions $\alpha \leq \beta$, i.e.,
\[
\Box_\alpha \phi \rightarrow \Box_\beta \phi.
\]
This criterion allows us to reformulate the notion of consistency not as a matter of probability amplitudes, but in terms of structural compatibility across regimes. The result is a non-epistemic reconstruction of quantum branching grounded in modal stability rather than in metaphysical duplication of realities.

\begin{itemize}
    \item Decoherence $\sim$ dynamic stabilization across layers;
    \item Histories $\sim$ consistent paths through stratified admissibility;
    \item Branching $\sim$ modal divergence without ontic multiplication.
\end{itemize}

\subsection{Modal Ontology and Constraint Structures}

The stratified semantics proposed in this logic does not merely refine modal distinctions; it redefines the ontological basis of modality. Rather than interpreting modal operators as quantification over pre-existing possibilities (as in standard Kripkean semantics), we treat modality as the articulation of structurally admissible regimes. Each stratum $\alpha \in I$ corresponds to a level of ontological constraint --- a domain in which certain propositions become dynamically stable due to underlying regularities.

In this framework, \emph{modal truth} is not absolute, but relative to constraint structure. A proposition $\phi$ is necessary at level $\alpha$ ($\Box_\alpha \phi$) if it holds under all transitions within the admissibility regime defined by $\alpha$. More precisely, truth conditions are stratified:
\[
\mathcal{M}, w \models \Box_\alpha \phi \quad \text{iff} \quad \forall v \in R_\alpha(w),\; \mathcal{M}, v \models \phi,
\]
where $R_\alpha$ respects both the stratification $\leq$ and internal stability conditions (e.g., coherence, conservation, or dynamical closure). The frame structure $F = (W, \leq, \{R_\alpha\}_{\alpha \in I})$ encodes the set of admissible transitions compatible with such constraints.

This view aligns modality with the theory of regimes of constraint, often discussed in contemporary philosophy of science and structural realism \cite{Knox2013, LadymanRoss2007, FriggHartmann2020}. In this setting, modal operators function not as metaphysical commitments to possible worlds, but as logical reflections of physical and theoretical admissibility.

Thus, modal validity becomes regime-relative:
\[
\text{If } \phi \text{ is valid at level } \alpha, \text{ then } \phi \text{ is coherent within } \mathcal{C}_\alpha,
\]
where $\mathcal{C}_\alpha$ denotes the set of constraint conditions defining the layer $\alpha$. This provides a foundation for interpreting modal logic as a logic of structured realization rather than one of global metaphysical possibility.

\section{Philosophical Discussion}
\subsection{Against Global Possibility Spaces}

Standard modal logic, particularly in the Lewisian framework, models possibility through a global space of possible worlds \cite{lewis1986}. In this paradigm, possibility and necessity are defined by relational accessibility across a fixed, totalized ontology of alternative realities. While powerful for analyzing counterfactuals, this model faces deep ontological and metaphysical problems.

First, it reifies the total space of possibilities as a completed structure, effectively collapsing the open-ended character of actualization. Actualization is not the selection of one world among many; it is the structured emergence of stability through local constraints. Possibility, in this framework, is not a pre-given range but a dynamically admissible neighborhood governed by stratified regimes \cite{fine2005, leihan2015}.

Second, this global model ignores the asymmetry and locality of real becoming. Temporal and physical processes do not unfold over a static space of alternatives but generate their own modal structures as they proceed. The Stratified Actualization Logic (SAL) proposed here replaces the global world-space with local layers of intensifying admissibility. Each layer $\alpha$ corresponds to a regime of coherence within which modal operators express admissibility, not mere metaphysical possibility.

This view aligns with structuralist interpretations of modality that regard modal truths as emerging from the internal consistency of physical or conceptual structures \cite{esfeld2017}. Rather than positing a metaphysically robust multiverse, we treat modal space as a layered structure of constraint compatibility, embedded within the process of actualization itself.

\subsection{Ontological Coherence and Logic}

Stratified Actualization Logic (SAL) proposes a shift from a metaphysical conception of modality—as abstract possibility—to a structural one, grounded in ontological coherence. In this framework, modal operators do not quantify over a pre-given space of alternatives but encode regimes of admissibility determined by layers of intensifying stability. Logical validity is no longer defined relative to a static semantic background but emerges from the coherence of localized evaluative structures.

This approach resonates with Fine’s critique of standard Kripke semantics, where he argues that modal truth should be grounded in truthmakers rather than possible-world quantification \cite{fine2012}. In his truthmaker semantics, necessity derives from what must be the case given the ontological structure of the actual world. Similarly, SAL locates necessity and possibility within the stability conditions internal to a system, rather than across a realm of alternatives.

Furthermore, the stratified model connects naturally to Schaffer’s proposal of grounding hierarchies \cite{schaffer2009}. In Schaffer’s account, metaphysical dependence is layered, and fundamental entities serve as the grounding base of derivative facts. Our modal stratification reflects a comparable hierarchy: each admissibility layer $\alpha$ can be seen as grounded in more fundamental constraints than layers $\beta > \alpha$, preserving coherence through intensification.

Modal logic, under this reformulation, becomes a logic of structural actualizability: a formal system encoding the compatibility and stabilization regimes within a layered ontology. This replaces the global semantics of accessibility with a local and processual account of coherence, aligning logical structure with ontological constraints rather than abstract metaphysical space. These concerns have been reinforced in recent literature on modal metaphysics. \citet{berto2019} develops a logic of open futures, rejecting the static ontology underlying classical Kripkean models. \citet{correia2021} argue for a dynamic understanding of grounding and truthmakers, aligning with the view that modal structure tracks ontological processes rather than pre-fixed domains.

\section{Conclusion and Further Work}

We have proposed a reformulation of modal logic as a stratified logic of actualization, grounded not in global possibility spaces but in layers of ontological admissibility. The formalism of Stratified Actualization Logic (SAL) replaces the static semantics of possible worlds with a dynamic structure indexed by intensity levels, representing degrees of stability, constraint, or causal coherence.

By introducing localized modal operators indexed to layers of admissibility, SAL captures the structural asymmetry and processual nature of actualization. This approach preserves logical consistency while accommodating phenomena such as temporal becoming, quantum stabilization, and ontological dependence, all of which elude representation in standard Kripkean models.

Further work may proceed along several directions. First, the integration of causal structures into SAL may yield a more complete account of dynamic constraint propagation, allowing a unified framework for modeling both modal and causal dependence. Second, categorical semantics—especially in the context of sheaf-theoretic or fibred logics—could offer a powerful formal underpinning to stratified evaluations across varying regimes. Finally, extending SAL toward intensional dynamics, in which transitions between layers encode transformations of potentiality, could link the logic of actualization with physical models of phase space evolution, decoherence, or even neural stabilization patterns.

\vspace{1em}

\paragraph{Limitations and Open Questions.}
While the proposed framework of stratified actualization provides a refined alternative to classical possible-worlds semantics, several important limitations remain. First, the semantic structure assumes a distributive base logic; its adaptation to non-distributive logics, such as relevance or substructural logics, remains an open direction. In contexts where classical conjunction and disjunction fail to distribute, the stratified accessibility relations \(\mathcal{R}_\alpha\) may no longer preserve modal coherence.

Second, although the framework allows for temporal interpretation via layered succession, it has not yet been integrated with formal systems like Computation Tree Logic (CTL) or CTL*. The challenge is to articulate stratified actualization within a branching-time formalism, preserving the asymmetry and locality of actualization while allowing for computational paths and temporal quantification. A precise translation or extension into these temporal systems remains to be developed.

The stratified model of modality proposed here thus opens new paths for representing ontological coherence, not as a property of world-selection, but as a dynamically maintained structure over constraint-governed systems.

\appendix

\section{Appendix: Formal Properties of SAL}

\subsection{A.1. Stratified Models and Local Evaluation}

We recall that a \textbf{stratified frame} is a tuple:
\[
\mathcal{F} = (W, \leq, \{\mathcal{R}_\alpha\}_{\alpha \in I})
\]
where:
\begin{itemize}
    \item $W$ is a nonempty set of worlds.
    \item $(I, \leq)$ is a partially ordered set of admissibility indices.
    \item For each $\alpha \in I$, $\mathcal{R}_\alpha \subseteq W \times W$ is a binary relation representing local accessibility at level $\alpha$.
\end{itemize}

A \textbf{stratified model} is a pair $\mathcal{M} = (\mathcal{F}, V)$ where $V : \text{Prop} \to \mathcal{P}(W)$ is a standard valuation function.

\paragraph{Truth conditions.} The truth of a formula $\phi$ at world $w \in W$, under index $\alpha$, is defined inductively:
\begin{align*}
\mathcal{M}, w \models p &\iff w \in V(p), \quad \text{for atomic } p \in \text{Prop} \\
\mathcal{M}, w \models \neg \phi &\iff \mathcal{M}, w \not\models \phi \\
\mathcal{M}, w \models \phi \wedge \psi &\iff \mathcal{M}, w \models \phi \text{ and } \mathcal{M}, w \models \psi \\
\mathcal{M}, w \models \Box_\alpha \phi &\iff \forall v \ (w \mathcal{R}_\alpha v \Rightarrow \mathcal{M}, v \models \phi) \\
\mathcal{M}, w \models \Diamond_\alpha \phi &\iff \exists v \ (w \mathcal{R}_\alpha v \text{ and } \mathcal{M}, v \models \phi)
\end{align*}

\subsection{A.2. Soundness}

We define the system \textbf{SAL} with the following axioms and rules:

\begin{itemize}
    \item All propositional tautologies.
    \item $\Box_\alpha (\phi \rightarrow \psi) \rightarrow (\Box_\alpha \phi \rightarrow \Box_\alpha \psi)$
    \item $\Box_\alpha \phi \rightarrow \Box_\beta \phi$ for all $\alpha \leq \beta$
    \item $\Box_\alpha \phi \rightarrow \phi$ if $\alpha$ is stable (i.e., designated in $I$)
    \item $\Diamond_\alpha \phi \rightarrow \Diamond_\beta \phi$ for all $\alpha \leq \beta$
    \item Modus ponens: from $\phi$, $\phi \rightarrow \psi$, infer $\psi$
    \item Necessitation (restricted): from $\phi$, infer $\Box_\alpha \phi$ only if $\alpha$ stable
\end{itemize}

\textbf{Theorem (Soundness).} If $\vdash_{SAL} \phi$, then $\mathcal{M}, w \models \phi$ for all stratified models $\mathcal{M}$ and worlds $w$.

\textit{Sketch.} The axioms reflect semantic conditions over $\mathcal{R}_\alpha$, stratification monotonicity $\alpha \leq \beta \Rightarrow \mathcal{R}_\alpha \subseteq \mathcal{R}_\beta$, and stability. Proof is by standard induction on derivation depth.

\subsection{A.3. Completeness}

Define a canonical model $\mathcal{M}^c = (W^c, \leq, \{R^c_\alpha\}, V^c)$ as follows:

\begin{itemize}
    \item $W^c = \{ \Gamma \subseteq \mathcal{L}_{SAL} \mid \Gamma$ is a maximal SAL-consistent set$\}$
    \item $\Gamma R^c_\alpha \Delta \iff \forall \phi, \Box_\alpha \phi \in \Gamma \Rightarrow \phi \in \Delta$
    \item $V^c(p) = \{ \Gamma \in W^c \mid p \in \Gamma \}$
\end{itemize}

\textbf{Lemma (Truth Lemma).} $\mathcal{M}^c, \Gamma \models \phi \iff \phi \in \Gamma$

\textbf{Theorem (Completeness).} If $\models \phi$, then $\vdash_{SAL} \phi$.

\textit{Sketch.} Follows from the truth lemma and construction of the canonical stratified model. The stratified accessibility is built respecting monotonicity and stability axioms.

\subsection{A.4. Complexity and Decidability}

While a complete analysis is reserved for future work, the system SAL with finite index set $(I, \leq)$ and finite branching assumptions inherits decidability from classical modal logic fragments.

\paragraph{Open question.} Under what topological or algebraic constraints on $(I, \leq)$ does SAL remain decidable?

\bibliographystyle{apsrev4-2}

\end{document}